 \documentclass[a4paper,10pt]{article}
 \usepackage{textcomp} %for degree C
 \usepackage{amsmath}
 \usepackage{graphicx}
 \usepackage[authoryear]{natbib}
 \usepackage{geometry}
 \usepackage{setspace}
 
\usepackage{hyperref}
 \usepackage{xcolor}
\hypersetup{
    colorlinks,
    linkcolor={red!50!black},
    citecolor={blue!50!black},
    urlcolor={blue!80!black}
}

\begin{document}
\title{How self-regulation, the storage effect and their interaction contribute
to coexistence in stochastic and seasonal environments}
\date{ }
\author{Coralie Picoche$^{*,1}$, Fr\'ed\'eric Barraquand$^{1,2}$}
%\titlerunning{Coexistence in seasonal environments}
%\authorrunning{C.Picoche \& F.Barraquand}
% \institute{{*}\email{coralie.picoche@u-bordeaux.fr}\\
% $^{1}$ University of
% Bordeaux, Integrative and Theoretical Ecology, LabEx COTE, B\^at. B2
% - All\'ee Geoffroy St-Hilaire, 33615 Pessac, France \\
% $^{2}$ CNRS, Institute
% of Mathematics of Bordeaux, 351 Cours de la Lib\'eration, 33405 Talence,
% France}
% \journalname{Theoretical Ecology}

\maketitle
\thispagestyle{empty} 
\textsuperscript{1}~{University of
Bordeaux, Integrative and Theoretical Ecology, LabEx COTE, France\bigskip{}
}{\large\par}

\textsuperscript{2}~{CNRS, Institute
of Mathematics of Bordeaux, France\bigskip{}}{\large\par}

% {\large{}\bigskip{}
% }{\large\par}
% 
% {\large{}\bigskip{}
% }{\large\par}
% 
% {\large{}\bigskip{}
% }{\large\par}
% 
% {\large{}\bigskip{}
% }{\large\par}

{*} Corresponding author. Email: coralie.picoche@u-bordeaux.fr
 
 \doublespacing

%\pagebreak{}

\begin{abstract}
Explaining coexistence in species-rich communities of primary producers
remains a challenge for ecologists because of their likely competition
for shared resources. Following Hutchinson's seminal
suggestion, many theoreticians have tried to create diversity through
a fluctuating environment, which impairs or slows down competitive
exclusion. However, fluctuating-environment models often only produce
a dozen of coexisting species at best. Here, we investigate how to
create richer communities in fluctuating environments, using an empirically
parameterized model. Building on the forced Lotka-Volterra model of
Scranton and Vasseur (Theor Ecol 9(3):353--363, 2016), inspired by phytoplankton communities,
we have investigated the effect of two coexistence mechanisms, namely
the storage effect and higher intra- than interspecific competition
strengths (i.e., strong self-regulation). We tuned the intra/inter
competition ratio based on empirical analyses, in which self-regulation
dominates interspecific interactions. Although a strong self-regulation
maintained more species (50\%) than the storage effect (25\%), we
show that none of the two coexistence mechanisms considered could
ensure the coexistence of all species alone. Realistic seasonal environments
only aggravated that picture, as they decreased persistence relative
to a random environment. However, strong self-regulation and the storage
effect combined superadditively so that all species could persist
with both mechanisms at work. Our results suggest that combining different
coexistence mechanisms into community models might be more fruitful
than trying to find which mechanism best explains diversity. We additionally
highlight that while biomass-trait distributions provide some clues
regarding coexistence mechanisms, they cannot indicate unequivocally
which mechanisms are at play.

%\textbf{Number of words: 239}
\end{abstract}

 Published in Theoretical Ecology (2019) \verb|doi:10.1007/s12080-019-0420-9|
\pagebreak{}

\section{Introduction}

The continued maintenance of diversity in spite of widespread competition
has long bewildered ecologists, especially for primary producers such
as phytoplankton that share the same basic resources \citep{hutchinson_paradox_1961}.
A solution for the `paradox of the plankon' was proposed by Hutchinson:
temporal variation of the environment. By making the identity of the
fittest species change over time, temporal variation could render
competitive exclusion less likely, which was confirmed by early experiments
\citep{sommer_paradox_1984}. However, it has been shown later that
inclusion of temporal variability \emph{per se} in competition models
is not sufficient for maintaining a realistic diversity \citep{chesson_roles_1997,fox_intermediate_2013}.
Additional mechanisms such as the storage effect \citep{chesson_multispecies_1994,ellner_how_2016}
or a relative nonlinearity of competition \citep{armstrong_r.a._competitive_1980,chesson_mechanisms_2000,descamps-julien_stable_2005,jiang_temperature_2007,fox_intermediate_2013}
need to be introduced for diversity to maintain. Moreover, richness
rarely exceeds a handful to a dozen of species in modeled competitive
communities in fluctuating environments, except when external inputs
from immigration sustain diversity \citep[e.g., ][]{huisman_towards_2001,jabot_non-random_2016}.
To our knowledge, the effect of temporal variability on persistence
in competition models has mostly been examined in theoretical communities
of 2 to 3 species \citep[e.g.,][]{chesson_roles_1997,litchman_competition_2001,li_effects_2016,miller_evolutionary_2017}.

One of the richest modeled communities that we identified can be found
in \citet{scranton_coexistence_2016}, which is based on temperature
variation and different thermal optima for each species \citep{moisan_modelling_2002}.
In this model, the synchronizing effect of the environment and the
storage effect can maintain 12 phytoplankton-like species on average.
\citet{scranton_coexistence_2016} described daily temperature as
a random noise, i.e., independent and identically distributed Gaussian
random variates over time. However, under most latitudes, seasonality
drives part of the environmental variation: over short timescales,
random temporal variations often only add noise to a largely deterministic
seasonal trend \citep{scheffer_seasonal_1997,boyce_environmental_2017,barraquand2018coastal}.

Seasonal forcing of parameters can strongly affect the dynamics of
model communities by synchronizing species to the seasonal signal
or even promoting oscillations with lower frequency \citep{rinaldi_multiple_1993,barabas_community_2012,miller_evolutionary_2017,vesipa_impact_2017}.
How seasonality affects coexistence, as opposed to a randomly fluctuating
environment, is therefore a key feature of this paper. Our present
work can be seen as an attempt to blend \citet{scranton_coexistence_2016}'s
stochastic framework with the periodic environments of \citet{barabas_community_2012},
to better represent the mixture of stochastic and deterministic environmental
forces affecting phytoplankton community dynamics.

What other key features of field communities should be considered
when modeling phytoplankton? Strong self-regulation, with intraspecific
competition much stronger than interspecific interactions, has been
found to be widespread in terrestrial plant communities \citep{adler_competition_2018},
animal communities \citep{mutshinda_what_2009}, and phytoplanktonic
communities \citep{barraquand2018coastal}. We will therefore insert
those niche differences, manifesting as strong self-regulation, into
our models of phytoplankton competition. The interaction between environment
variability and niche overlap was investigated by \citet{abrams_niche_1976}
but his results did not extend to communities more diverse than 4
species; our objective is therefore to see how those mechanisms interact
for species-rich communities.

Niche models have often been opposed to the neutral theory \citep{hubbell_unified_2001},
where dispersal and drift can ensure a transient coexistence of many
species, but several authors have attempted to blend niche and neutral
processes \citep{gravel_reconciling_2006,scheffer_self-organized_2006,carmel_using_2017}.
An intriguing offshoot of these attempts is the concept of `clumpy
coexistence' \citep{scheffer_self-organized_2006}, whereby simultaneous
influences of both niche and neutral processes create several clumps
of similar species along a single trait axis. Niche differences enable
coexistence of multiple clumps \citep{chesson_mechanisms_2000} while
within-clump coexistence occurs through neutral processes. This `emergent
neutrality' within groups \citep{holt_emergent_2006} has been proposed
as a unifying concept for niche and neutral theories (even though
the neutrality of the original model has been disputed due to hidden
niches, \citealp{barabas_emergent_2013}). Since then, clumpy coexistence
has been shown to occur in theoretical models incorporating a temporally
variable environment interacting with a thermal preference trait axis
\citep{scranton_coexistence_2016,sakavara_lumpy_2018}. The relationship
(or absence thereof) between biomass-trait distributions and coexistence
mechanisms is currently debated \citep{dandrea_challenges_2016},
although there are suggestions that clustering on trait axes under
competition may be a robust find \citep{dandrea_translucent_2018,dandrea_generalizing_2019}.

Here, we try to establish what are the relative contributions to coexistence
of the storage effect vs strong self-regulation, in a phytoplankton-like
theoretical community model with a large number of species. This led
us to cross combinations of seasonality vs randomness in the forcing
signal, presence of the storage effect or not, and intra- vs interspecific
competition intensity, in order to disentangle the contributions of
these factors to biodiversity maintenance and their potential interactions.
Alongside the resulting species richness, we also report which biomass-trait
distribution can be expected under a given combination of processes
leading to coexistence.

\pagebreak
\section{Methods}

\subsection*{\emph{Models}}

The model described in \citet{scranton_coexistence_2016} is based
on the Lotka-Volterra competition model. Fluctuations in the environment
are introduced in the model by temperature-dependent intrinsic growth
rates (see Eq. \ref{eq:modelLV}-\ref{eq:growth}, all coefficients
are defined in Table~\ref{tab:Coefficients-}) so that the community
dynamics can be expressed as:

\begin{eqnarray}
&\frac{dN_{i}}{dt}  = & r_{i}(\tau)N_{i}\left(1-\sum_{j=1}^{S}\alpha_{ij}N_{j}\right)-mN_{i}\label{eq:modelLV}\\
&r_{i}(\tau)  = & a_{r}(\tau_{0})e^{E_{r}\frac{(\tau-\tau_{\text{0}})}{k\tau\tau_{0}}}f_{i}(\tau)\label{eq:growth}\\
\textnormal{where } & f_{i}(\tau)  = & \begin{cases}
e^{-|\tau-\tau_{i}^{opt}|^{3}/b_{i}}, & \tau\leq\tau_{i}^{opt}\\
e^{-5|\tau-\tau_{i}^{opt}|^{3}/b_{i}}, & \tau>\tau_{i}^{opt}
\end{cases}\label{eq:eppley_curve}\\
\textnormal{and} & b_{i}\text{ is defined by numerically solving} & \int r_{i}(\tau)d\tau=A\label{eq:niche_breadth}
\end{eqnarray}

Model parameters are detailed in Table~\ref{tab:Coefficients-}, and
we set their values to match the features of phytoplankton communities
as in Scranton and Vasseur's work \citeyearpar{scranton_coexistence_2016}.
The niche of each species is defined by its thermal optimum $\tau_{i}^{opt}$.
Thermal performance curves defined in Eq. \ref{eq:eppley_curve} are
parameterized so that all species share the same niche area (Eq. \ref{eq:niche_breadth}),
which sets a trade-off between maximum growth rates and niche width.

\begin{table}[!ht]
\caption{Parameter definitions and values for the model described in Eqs. \ref{eq:modelLV}-\ref{eq:niche_breadth}.
Parameter values are not specified when they vary over time and/or
with the species considered. \label{tab:Coefficients-}}

\centering{}%
\begin{tabular}{ccc}
\hline
Name & Definition & Value (unit)\\
\hline
$S$ & Initial number of species & 60 (NA)\\
$N_{i}$ & Biomass density of the $i^{th}$ species & (kg/area)\\
$\tau$ & Temperature & (K)\\
$r_{i}(\tau)$ & Growth rate of species $i$ as a function of temperature & $(\frac{\text{kg}}{\text{kg\ensuremath{\times}year}}$)\\
$\alpha$ & Baseline strength of competition & 0.001 (area/kg)\\
$b_{i}$ & Normalization constant for the thermal decay rate & $(K^{3}$)\\
$m$ & Mortality rate & $15(\frac{\text{kg}}{\text{kg\ensuremath{\times}year}}$)\\
$\tau_{0}$ & Reference temperature & 293 (K) / 20 (\textdegree C)\\
$a_{r}(\tau_{0})$ & Growth rate at reference temperature & $386(\frac{\text{kg}}{\text{kg\ensuremath{\times}year}}$)\\
$E_{r}$ & Activation energy & 0.467 (eV)\\
$k$ & Boltzmann's constant & $8.6173324.10^{-5}(\text{eV.K}^{-1})$\\
$f_{i}(\tau)$ & Fraction of the maximum rate achieved for the $i^{th}$ species & (NA)\\
$\mu_{\tau}$ & Mean temperature & 293 (K)\\
$\sigma_{\tau}$ & Standard deviation for temperature & 5 (K)\\
$\tau_{\text{min}}$ & Minimum thermal optimum & 288 (K)\\
$\tau_{\text{max}}$ & Maximum thermal optimum & 298 (K)\\
A & Niche breadth & $10^{3.1}(\frac{\text{kg}}{\text{kg\ensuremath{\times}year}})$\\
$\tau_{i}^{\text{opt}}$ & Thermal optimum for growth of the $i^{th}$ species & (K)\\
$\theta$ & Scaling between random and seasonal noise & (0;1.3) (NA)\\
$\kappa$ & Ratio of intra-to-interspecific competition strength & (1;10) (NA)\\
\hline
\end{tabular}
\end{table}

The original environmental forcing is a normally distributed variable
centered on 293 K (20\textdegree C), with a 5 K dispersion. Temperature varies
from one day to the next, but is kept constant throughout the day.
At and above the daily scale, temperature could therefore be considered
as a white noise \citep{vasseur_color_2004}. However, from a mathematical
viewpoint, the noise is slightly autocorrelated as the integration
process goes below the daily time step. We therefore use the expression
`random noise' to describe this forcing, as opposed to the `seasonal
noise' described hereafter. To construct the seasonal noise, we add
to the random forcing signal a lower-frequency component, using a
sinusoidal function with a period of 365 days (Eq. \ref{eq:Seasonal_signal}).
We tune the ratio of low-to-high frequency with the variable $\theta$
so as to keep the same energy content - i.e., equal total variance
- in the forcing signal.

\begin{equation}
\tau(t)=\mu_{\tau}+\theta\sigma_{\tau}\sin\left(2\pi t\right)+\epsilon_{t},\text{where }\epsilon_{t}\sim\mathcal{N}\left(0,\sigma_{\tau}\sqrt{1-\frac{\theta^{2}}{2}}\right)\label{eq:Seasonal_signal}
\end{equation}

Note that the upper limit for $\theta$, $\sqrt{2}$, corresponds
to a completely deterministic model which we do not explore here (but
see \citet{zhao1991qualitative} proving bounded coexistence). We
choose to keep the stochasticity in the signal and to model a plausible
temperature signal with $\theta=1.3$ (illustrated in Fig.~\ref{fig:Times-series_temperature_species}b)
when considering a seasonal forcing of the dynamics.

The formulation of the forced Lotka-Volterra model of \citet{scranton_coexistence_2016}
implies a storage effect, as the net effect of competition exerted
by species $j$ on $i$ is the product of the temperature-related
growth rate $r_{i}(\tau)$ and the competitive strength $\alpha_{ij}$
exerted by species $j$ multiplied by its abundance $N_{j}$. Therefore,
total net competition ($\sum_{j=1}^{S}r_{i}(\tau)\alpha_{ij}N_{j}$)
covaries positively with the growth rate values $r_{i}(\tau)$, which
defines the storage effect \citep{chesson_multispecies_1994,fox_intermediate_2013,ellner_how_2016}.
To remove the assumption of an explicit storage effect, we create
another version of the model using the mean value of a species' growth
rate ($\bar{r_{i}}$) to weight the interaction coefficients (see
Eq. \ref{eq:eqLV_noforcecompetition}). The mean growth rate value
is computed by first generating the temperature time series and then
averaging all $r_{i}$ over the corresponding sequences of $\tau$
values.

\begin{equation}
\frac{dN_{i}}{dt}=N_{i}\left(r_{i}(\tau)-\sum_{j=1}^{S}\bar{r}_{i}\alpha_{ij}N_{j}\right)-mN_{i}\label{eq:eqLV_noforcecompetition}
\end{equation}

Following Eq. \ref{eq:eqLV_noforcecompetition}, net competition remains
unaffected by the environmental conditions, in contrast to intrinsic
growth rates, while preserving the same average magnitude of competition
as in Eq. \ref{eq:modelLV}.

Strong self-regulation is ensured by the addition of the coefficient
$\kappa$, which is the ratio of intra-to-interspecific competition
strength. We can therefore re-write the interaction coefficients $\alpha_{ij}$
in Eq. \ref{eq:alpha_def}

\begin{equation}
\alpha_{ij}=\alpha\left(1+(\kappa-1)\delta_{ij}\right)\label{eq:alpha_def}
\end{equation}

where $\delta_{ij}$ is the Kronecker symbol, equal to 1 if $i=j$
and to 0 otherwise. The value of the parameter $\kappa=10$ was chosen
from analyses of phytoplanktonic data \citep{barraquand2018coastal}. The values of non-zero interspecific competition coefficients reported in Fig.~5 of \citet{barraquand2018coastal} are somewhat higher than here (and $\kappa$ lower, closer to 4) because the best-fitting model actually sets to zero some intergroup competition coefficients; using a model where all $\alpha_{ij}$ are non-zero leads in contrast to $\kappa=10$). Hereafter, the expression ``strong self-regulation'' characterizes
dynamics where the intraspecific competition strength is 10 times
higher than the interspecific competition strength, as opposed to
``equal competitive strengths'' where intra- and interspecific competition
strengths are equal.

In addition to two types of environmental forcings (random noise with
$\theta=0$, and seasonal noise with $\theta=1.3$), we compare the
results for four versions of the original model: with and without
an explicit storage effect (Eq. \ref{eq:modelLV} and Eq. \ref{eq:eqLV_noforcecompetition},
respectively); with strong self-regulation or equal intra- and inter-competition
strength ($\kappa=10$ or $\kappa=1$, respectively). These are summed
up in Table~\ref{tab:ModelsOfGrowthRates}.
\begin{center}
\begin{table}[!ht]
\begin{centering}
\begin{small}
%\hspace{-50pt}
\begin{tabular}{ccc}
\hline 
$\frac{1}{N_{i}}\frac{dN_{i}}{dt}+m$ & Storage effect & No storage effect\\
\hline 
Strong self-regulation ($\kappa=10$) & $r_{i}(\tau)\left(1-\sum_{j=1}^{S}\alpha\left(1+9\delta_{ij}\right)N_{j}\right)$ & $r_{i}(\tau)-\sum_{j=1}^{S}\bar{r}_{i}\alpha\left(1+9\delta_{ij}\right)N_{j}$\\
\hline 
Equal competitive strengths ($\kappa=1$) & $r_{i}(\tau)\left(1-\sum_{j=1}^{S}\alpha N_{j}\right)$ & $r_{i}(\tau)-\sum_{j=1}^{S}\bar{r}_{i}\alpha N_{j}$\\
\hline 
\end{tabular}
\end{small}
\par\end{centering}
\caption{\label{tab:ModelsOfGrowthRates}Growth rate of species $i$ in the
four models}
\end{table}
\par\end{center}

\subsection*{Set-up}

We replicate the `Species sorting' experiment of \citet{scranton_coexistence_2016}
so as to investigate how the structure of synthetic phytoplankton
communities varies under the different scenarios we described above.
We focus on the dynamics of a community initialized with 60 species
with thermal optima uniformly spaced along the interval {[}15\textdegree C, 25\textdegree C{]},
and with the same initial density $\left(\frac{\ensuremath{1}}{\alpha S}\right)$.
Each simulation is run for 5000 years in 1-day intervals. When the
density of a species drops below $10^{-6}$, it is considered extinct.
For each combination of parameters (type of environmental signal,
storage effect and self-regulation), we run 100 simulations.

All simulations are run with Matlab's ode45 algorithm, an adaptive
Runge-Kutta (4,5) integration scheme with an absolute error tolerance
of $10^{-8}$, and relative error tolerance of $10^{-3}$. The code
is available in a GitHub repository\footnote{\texttt{https://github.com/CoraliePicoche/Seasonality}}.

\pagebreak
\section{Results}

\begin{figure}[!ht]
\begin{centering}
\includegraphics[width=0.95\textwidth]{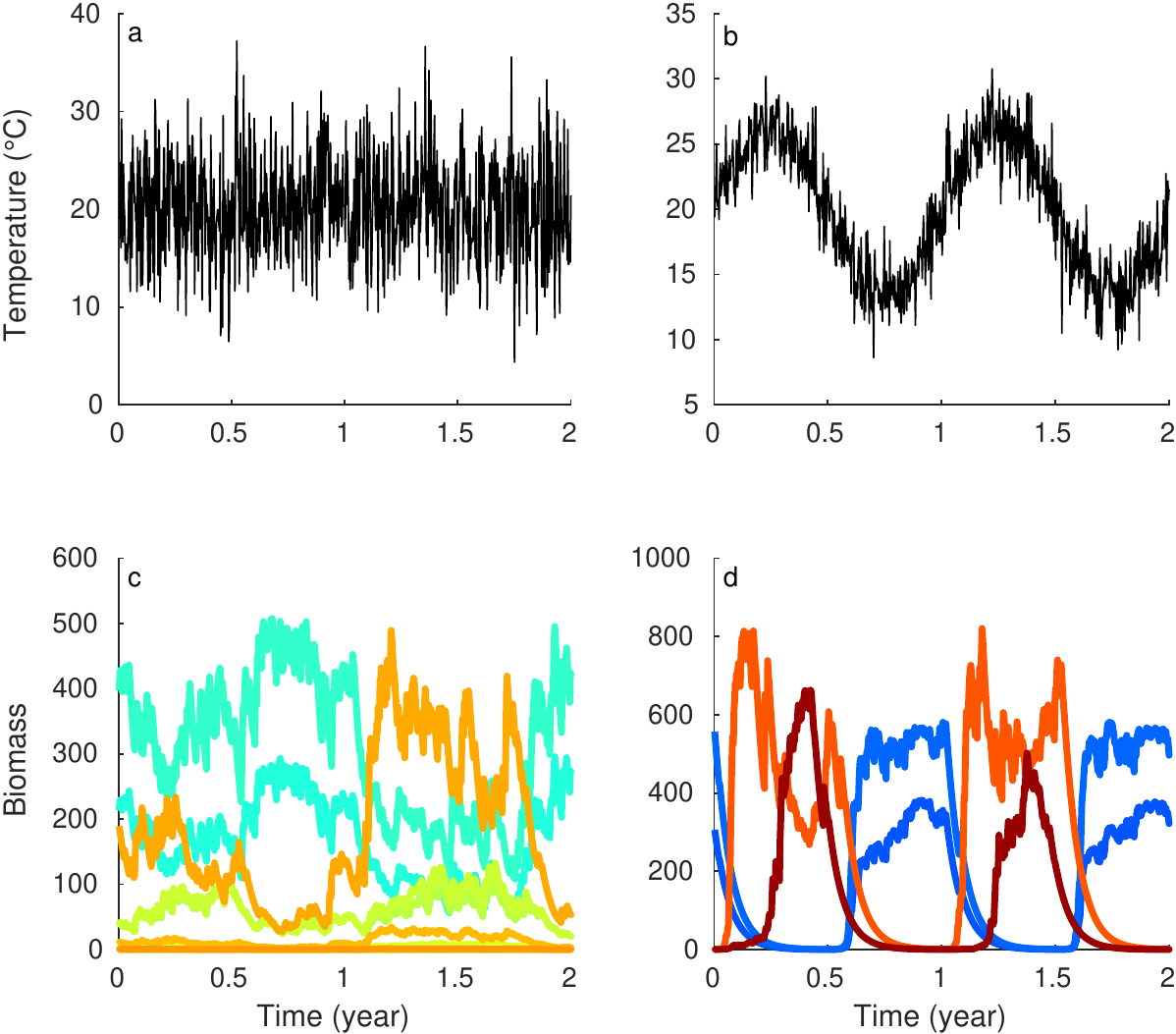}
\par\end{centering}
\caption{Time series of temperature (top; a-b) and extant species (bottom;
c-d) for the 2 last years of a 5000-year simulation, with storage
effect but no differences between intraspecific and interspecific
competition strengths. The forcing temperature is either a random
noise (a) or a seasonal noise (b), leading to community dynamics with
more erratic fluctuations (c) vs seasonally structured fluctuations
(d). Line colors of species biomasses correspond to their thermal
optimum (from blue, corresponding to low thermal optimum, to red,
corresponding to high thermal optimum).\label{fig:Times-series_temperature_species}}
\end{figure}

Typical dynamics of the community following Eq. \ref{eq:modelLV}
(the model of \citealp{scranton_coexistence_2016}), with both a purely
Gaussian noise (original choice of \citealp{scranton_coexistence_2016};
Fig.~\ref{fig:Times-series_temperature_species}a) and a seasonal
noise described in Eq. \ref{eq:Seasonal_signal} (our variant, Fig.
\ref{fig:Times-series_temperature_species}b), are shown in Fig.~\ref{fig:Times-series_temperature_species}c
and d, respectively. A sinusoidal forcing produces the strongly seasonally
structured dynamics that are typical of phytoplankton. Even though
only 5 species can be seen in Fig.~\ref{fig:Times-series_temperature_species}c,
there were 14 species still present at the end of the simulation forced
by a random noise, with large disparities in the range of their biomasses.
A third of the species kept a biomass above 10 kg/area (setting area
= 1 ha, with a depth of a few meters, produces realistic standing
biomasses; \citealp{reynolds2006ecology}) while 6 out of the 14 species
biomasses remained below the unit. All persisting species in the random
noise simulation were clustered within a 3.2\textdegree C-range of thermal optima
(see the biomass distribution as a function of the thermal optimum
in Electronic Supplementary Material, Fig.~A1). No obvious temporal
patterns (e.g., cycles) could be seen in the community forced by random
noise. On the contrary, seasonal cycles were clear in the seasonally-forced
case of Fig.~\ref{fig:Times-series_temperature_species}d. Only 4
species coexisted at the end of the simulation with seasonal noise,
gathered in two groups with large thermal optimum differences (5.7\textdegree C
between the maximum thermal optimum of the first group and the minimum
thermal optimum of the second group). When temperatures were high,
the group with higher thermal optima reached its maximum biomass,
then as temperature decreases through the season, these species leave
room for the growth of the low-temperature group.

\begin{figure}[!ht]
\begin{centering}
\includegraphics[width=0.95\textwidth]{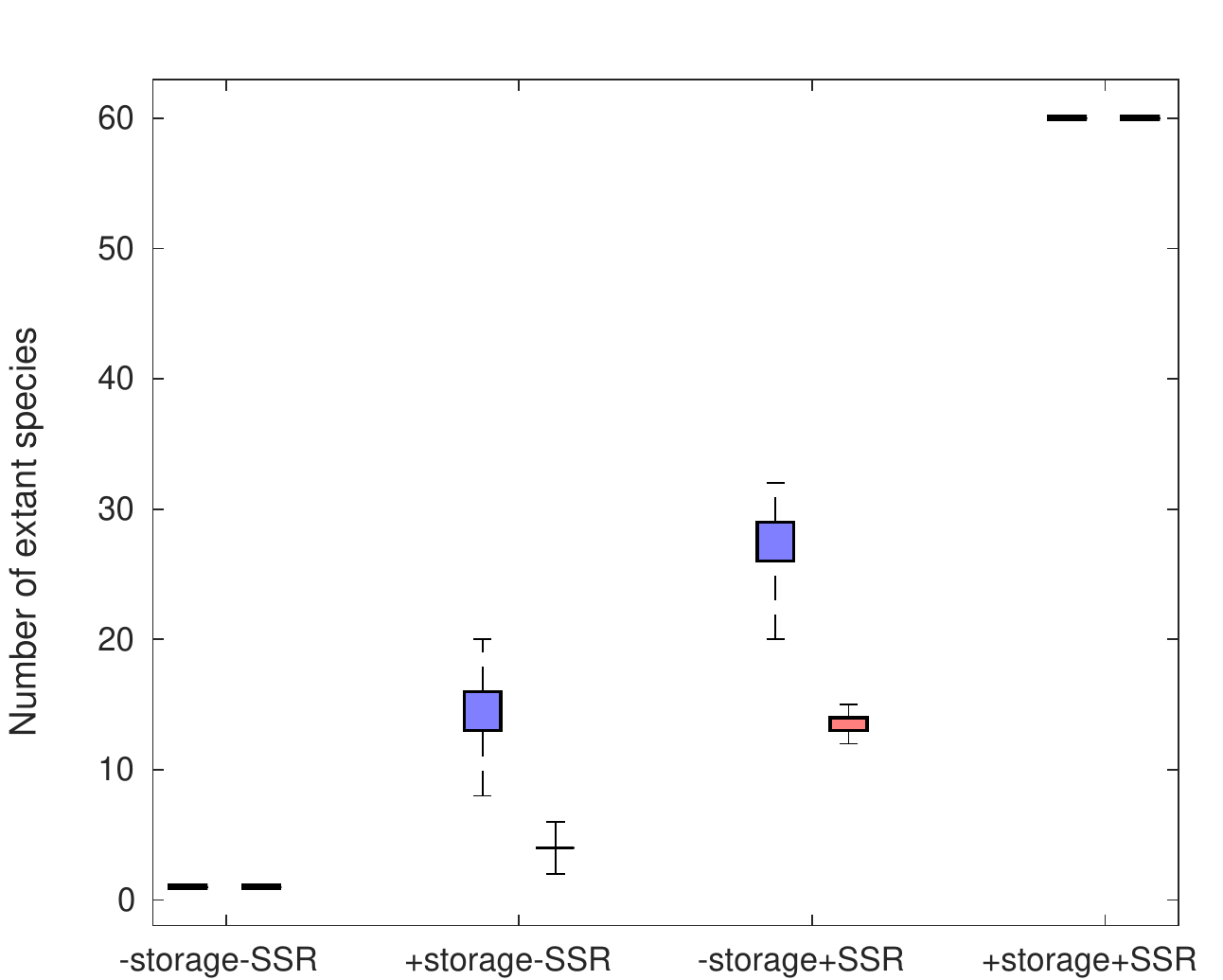}
\par\end{centering}
\caption{Number of species still present at the end of 100 simulations (5000
years each), initialized with 60 species, with a random (blue) or
a seasonal forcing signal (red). The signs + or -storage refer to
presence or absence of the storage effect, respectively; + / - SSR,
presence or absence of Strong Self-Regulation, respectively. Community
compositions are stable in the cases -storage-SSR and +storage+SSR,
for which 1 or 60 species are still present at the end of all simulations,
respectively. Due to low variance, the whiskers here represent min
and max rather than 1.5 interquartile range. \label{fig:Persistence-of-species}}
\end{figure}

The decrease in persistence due to seasonality observed in Fig.~\ref{fig:Times-series_temperature_species}
was confirmed in all our simulations (Fig.~\ref{fig:Persistence-of-species}).
In cases where final species richness varied from one simulation to
another (namely, the two middle cases in Fig.~\ref{fig:Persistence-of-species}:
with storage effect but without strong self-regulation, or without
storage effect but with strong self-regulation), seasonality reduced
the number of extant species to, on average, 27\% and 48\% of their
original values, respectively (Fig.~\ref{fig:Persistence-of-species}).
A seasonal signal therefore led to a much smaller average persistence.
There was also less variance in persistence between seasonally forced
simulations compared to random noise simulations.

Both a strong self-regulation and the storage effect markedly increased
persistence. Without any of these coexistence mechanisms, only one
species persisted at the end of the simulations. When only the storage
effect was present, the number of extant species varied between 8
and 20 (14.8 $\pm$ 2.4) with random noise, or 2 and 6 (4.1 $\pm$
0.7) with a seasonal signal. On the other hand, when only a strong
self-regulation was present, the number of extant species nearly doubled,
varying between 20 and 32 (27.5 $\pm$ 2.4), or 12 and 15 (13.3 $\pm$
0.6), with a random or a seasonal noise, respectively. Remarkably,
when the storage effect and a strong self-regulation both affected
the community dynamics, all species persisted in the community: the
number of species coexisting with both mechanisms present is greater
than the sum of the species coexisting with either mechanism alone.
The two mechanisms therefore combine superadditively, as their interaction
has a positive effect on the richness of the community.

\begin{figure}[!ht]
\begin{centering}
\includegraphics[bb=0bp 0bp 356bp 296bp,width=0.95\textwidth]{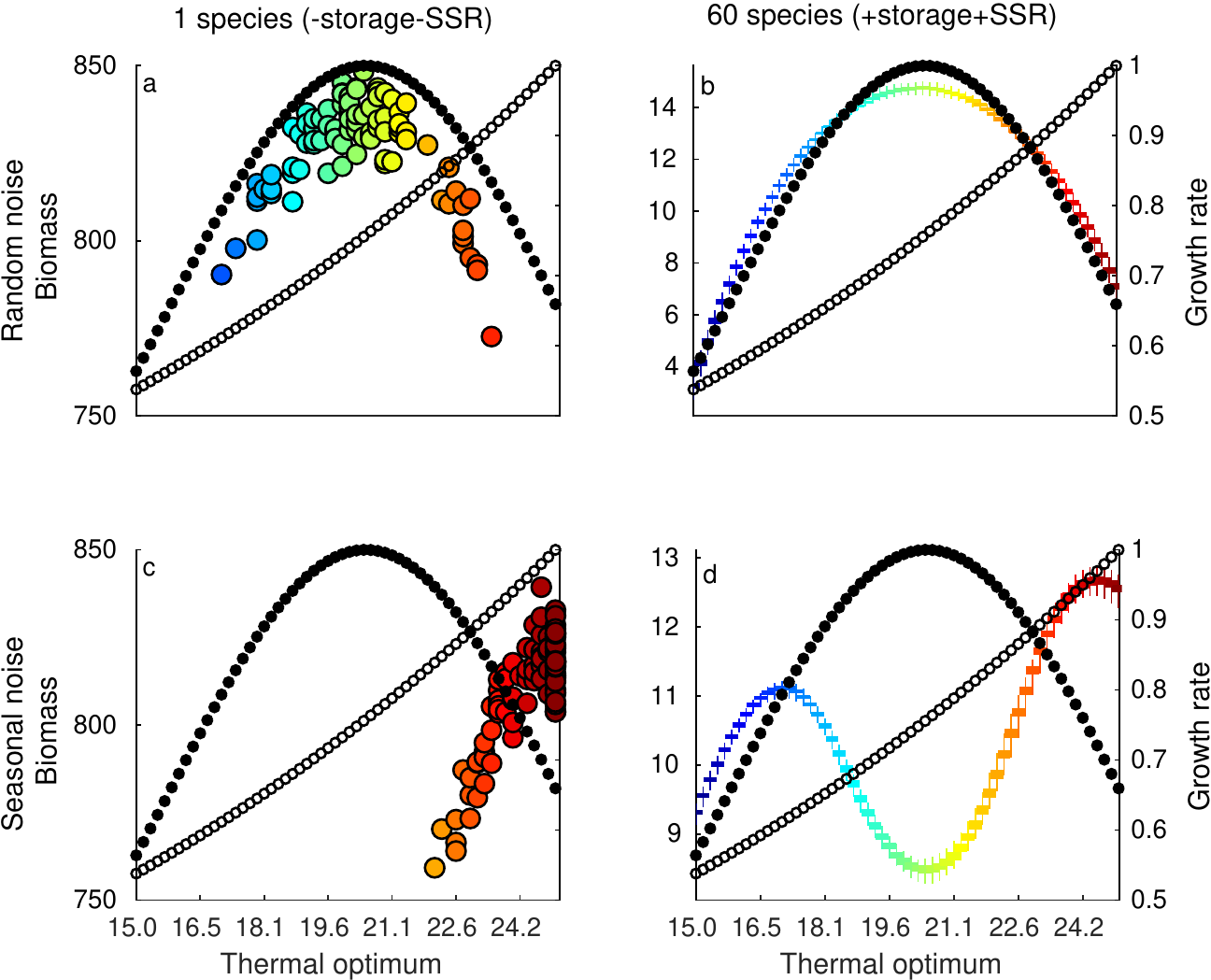}
\par\end{centering}
\caption{Mean biomass distribution over the last 200 years for 100 simulations,
as a function of the species thermal optima. Here we consider the
two stable-composition cases and two types of forcing signal. On the
left column, simulations without storage effect nor strong self-regulation
are presented. Only one species is present at the end of the simulations
and its mean value is represented by one large colored circle per
simulation. There can be several circles for the same species, corresponding
to multiple simulations ending with this species alone. On the right
column, simulations with storage effect and strong self-regulation
are represented. All species are present at the end of the simulations
and small boxplots present the variation in the temporal average of
biomass with a given trait, for 100 simulations. The forcing signal
is either a random (top) or a seasonal noise (bottom). Each species
is identified by its thermal optimum through its color code. Scaled
(divided by maximum) average and maximum growth rates are shown as
small filled and open circles, respectively, and are indexed on the
right y-axis. \label{fig:Mean-biomass-in_stable_cases}}
\end{figure}

The trait-biomass distribution of the community was affected by the
type of forcing even when the richness of the community was stable
(Fig.~\ref{fig:Mean-biomass-in_stable_cases}). Without storage effect
nor strong self-regulation, there was only one species left at the
end of the simulations. A random noise favored species with intermediate
thermal optima: the final species had a thermal optimum between 18.9\textdegree C
and 21.4\textdegree C (only a fourth of the initial range of thermal optima)
for two simulations out of three and the maximum final biomasses over
100 simulations was reached in this range (Fig.~\ref{fig:Mean-biomass-in_stable_cases}a).
This distribution may indicate a selection for the highest long-term
growth rates, averaged over time (see scaled growth rates in Fig.
\ref{fig:Mean-biomass-in_stable_cases}). Seasonality with no coexistence
mechanisms also led to a single final species but, in this case, the
species always had a higher maximum growth rate (thermal optimum above
22\textdegree C). Species with a higher thermal optimum were more likely to persist
and to reach a higher biomass at the end of the simulation. 38\% of
the simulations therefore ended with the species having the highest
temperature optimum, 25\textdegree C. The shift in trait distribution towards
higher maximum growth rates with a seasonal noise vs higher average
growth rates with a random noise was consistent for all model types
considered.

When both the storage effect and strong self-regulation were present,
the 60 initial species coexisted with almost no variation in their respective
biomasses from one simulation to the next (mean CV across simulations
is 0.008, averaged across species, Fig.~\ref{fig:Mean-biomass-in_stable_cases}b
and d). The forcing signal modified only the distribution of biomasses,
resulting in contrasted community structures despite equal richness.
With a random noise, the distribution was unimodal. On the contrary,
a seasonal signal led to a bimodal distribution (centered on 17.0\textdegree C
and 24.4\textdegree C), each corresponding to one season, with highest biomasses
for higher thermal optima (Fig.~\ref{fig:Mean-biomass-in_stable_cases}d).
The minimum biomass was reached for the highest long-term average
growth rate at an intermediate temperature (20.4\textdegree C).

\begin{figure}[!ht]
\begin{centering}
\includegraphics[bb=0bp 0bp 428bp 338bp,width=1\textwidth]{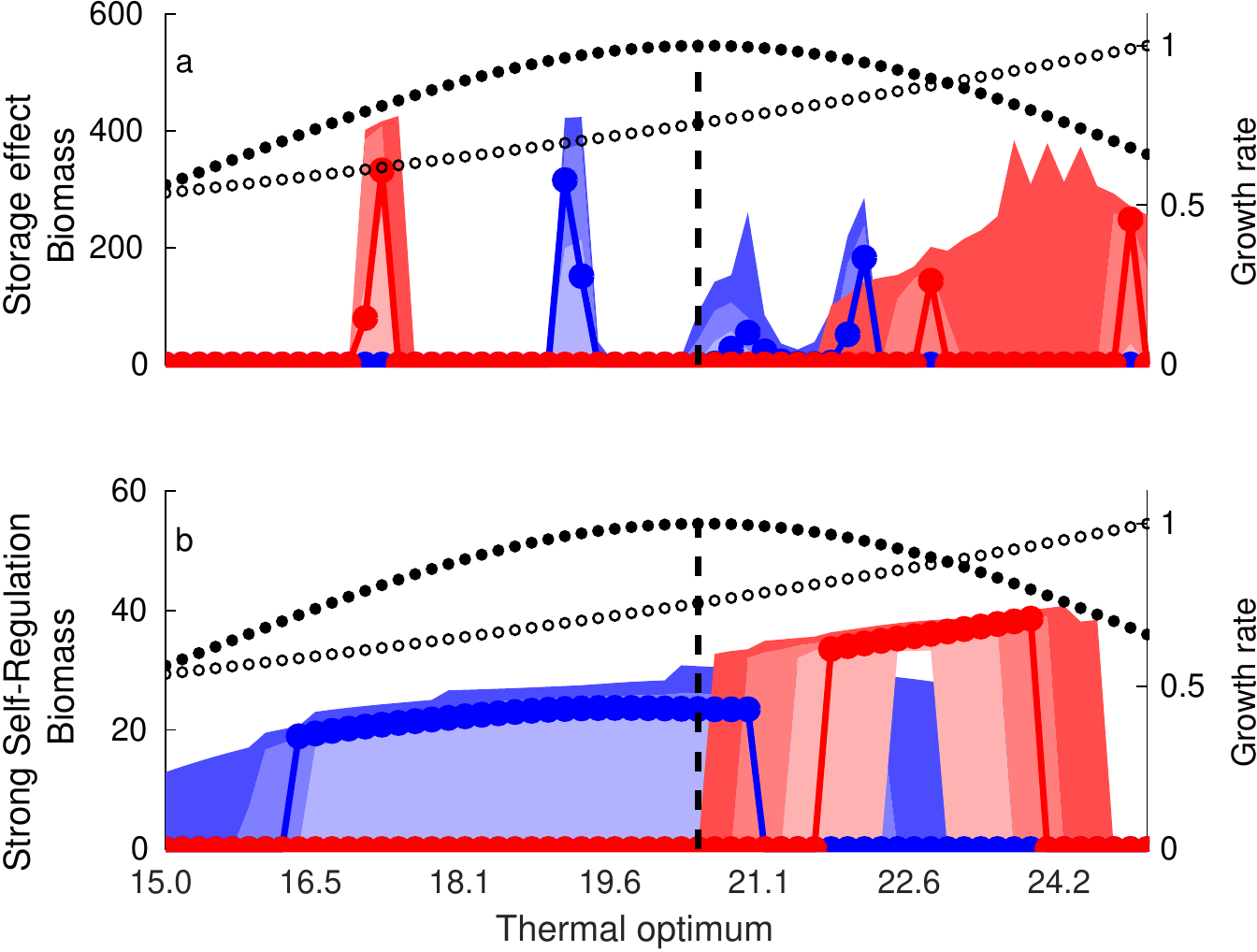}
\par\end{centering}
\caption{Mean biomass distribution over the last 200 years for 100 simulations,
as a function of thermal optima, (a) with storage effect and equal
competitive strengths and (b) without storage effect, with strong
self-regulation. The forcing signal is either a random (in blue) or
a seasonal noise (in red). Shades of the same color correspond to
the 50th, 90th and 100th percentiles of the distributions while colored
lines correspond to one representative simulation. Scaled (divided
by maximum) average and maximum growth rates are shown as filled and
open and circles, respectively, and indexed on the right y-axis. The
maximum average growth rate is indicated by the dashed line.\label{fig:Unstable_cases}}
\end{figure}

In cases where the richness of the community varied, the overall shape
(multimodal vs unimodal) of the marginal distribution of extant species
with respect to the trait axis were similar for both types of environmental
forcings (Fig.~\ref{fig:Unstable_cases}). By contrast, the type of
coexistence mechanism generated different shapes. Indeed, the storage
effect (when acting alone) led to a multimodal biomass distribution
with respect to thermal optima. We always observed 3 modes with a
random noise and 3 modes in 95\% of the seasonal simulations (Fig.
\ref{fig:Unstable_cases}a). With a random noise, extant species were
grouped in rather similar clumps regarding species thermal optima
(between 18.8\textdegree C and 22.2\textdegree C) whereas species tended to be further apart
in the seasonal case, covering a total range of 7.7\textdegree C, with species
grouping in the higher part of the thermal range, above 22\textdegree C. On the
other hand, strong self-regulation led to a quasi-uniform biomass
distribution (Fig.~\ref{fig:Unstable_cases} b). Species in communities
forced by a random noise stayed in the lower range of thermal optima
(in 96\% of the simulations, the highest thermal optimum was 22.4\textdegree C,
see Fig.~A2 in Electronic Supplementary Material) while they were
filtered out in communities subjected to a seasonal fluctuation of
their environment, for which species with thermal optima above 20.5\textdegree C
persisted. As before (Fig.~\ref{fig:Mean-biomass-in_stable_cases}
c,d), seasonality promoted species with a higher maximum growth rate,
since the autocorrelated temperatures enabled them to achieve this
highest growth rate for a longer period of time than a random noise
would have.

\pagebreak
\section{Discussion}

We have simulated competitive Lotka-Volterra dynamics forced by a
fluctuating temperature under a range of scenarios allowing more or
less coexistence. Two coexistence mechanisms, the storage effect and
strong self-regulation (i.e., intraspecific competition much stronger
than interspecific competition), could be either present or absent,
which led to four scenarios. These four scenarios were crossed with
two possibilities for the forcing signal, a random noise (mostly white)
and a stochastic yet seasonal signal, both with equal temporal variance.
Our investigation therefore built on the model of \citet{scranton_coexistence_2016},
which included a random forcing signal and a storage effect, but considered
seven additional combinations of mechanisms. This was motivated by
our wish to include two observed features of phytoplankton dynamics:
seasonal cycles \citep{winder_annual_2010,boyce_environmental_2017}
and strong self-regulation \citep{chesson_mechanisms_2000,adler_coexistence_2010,barraquand2018coastal}.
Many mechanisms can lead to intraspecific competition being stronger
than interspecific competition: nonlinearities in the functional forms
of competition or mutualism that contribute to increasing self-regulation
\citep{kawatsu2018density}, or predation as well as parasitism (see
e.g., the generalist predators in \citealp{haydon1994pivotal}). Strong
self-regulation seems an ubiquitous feature in competition networks
of primary producers \citep{adler_competition_2018}, and perhaps
even more general networks \citep{barabas_self-regulation_2017}.

Before discussing the ecological interpretation of our results, we
first recall some technical assumptions made in this study. All our
simulations lasted for a fixed duration (5000 timesteps) as in \citet{scranton_coexistence_2016}.
While short- and medium-term transients (a few years to hundreds of
years) are completely negligible at the end of the time series, very
long transients can remain in this class of models \citep{scheffer_self-organized_2006,hastings_transient_2018}:
these are not mere artefacts but instead traduce the fact that some
processes (e.g., exclusion of a species) can be really slow. We realized
that convergence was incomplete after 5000 years in some cases (e.g.,
random noise + storage effect + equal competitive strength). Such
simulations would take up to 15 000 years to converge and the rate
of convergence would slow over time. We could have considered longer
time intervals, but comparison with the values reported by \citet{scranton_coexistence_2016}
would then have been compromised. Another way to shorten the transients,
suggested by a referee (GB), is to vary the mortality parameter. This
did not alter the conclusions (see Appendix B in Electronic Supplementary
Material). Unfortunately, added variability also shifts the model
further away from neutral dynamics (when intra and interspecific competition
strengths are equal), which renders comparisons difficult. All things
considered, we therefore kept the 5000-year time window for integration.

Another strong assumption pertains to competition coefficients. To
allow for comparison with \citet{scranton_coexistence_2016}, we did
not introduce variability in intraspecific competition strength or
interspecific competition strength. By contrast, data-based coefficients
vary between species \citep{barraquand2018coastal}, with a majority
of weak interactions (as suggested in \citealp{wootton_measurement_2005})
and more variance in intraspecific coefficients. \citet{stump_multispecies_2017}
recently considered the potential effects of competition coefficient
variability (also called non-diffuse competition), as did \citet{kokkoris_variability_2002};
more variance in interspecific competition strength is usually detrimental
to coexistence for an equal amount of self-regulation (see \citet{stump_multispecies_2017}
for a classification of the various effects). Setting the competition
coefficients using a multidimensional trait-based framework, like
that of \citet{ashby_competing_2017}, would provide a natural development
to the work presented here; it is in our opinion difficult to speculate
on those variance effects because both intra- and interspecific competition
coefficient variances may matter to community persistence.

Finally, our study is limited to communities whose species have fast
population dynamics relative to the yearly timescale, like phytoplankton
and likely other fast-living organisms, so that many generations can
occur in a year. Persistence in community with slower dynamics may
be affected differently by seasonality \citep{miller_evolutionary_2017}.
This is especially true for species with generations that extend over
multiple years. In models where trophic interactions are implemented,
seasonality has been shown to promote multiyear cycles and the existence
of chaotic attractors \citep{rinaldi_multiple_1993,taylor_how_2013,tyson_seasonally_2016}.
These rich dynamics of consumers may feed back into the lower trophic
levels: \citet{dakos_interannual_2009} present a planktonic community
with seasonally-entrained chaotic dynamics which may be partly due
to zooplanktonic predation. Predation probably entails additional
niche differences, possibly with an emerging self-regulation created
by predation processes \citep{chesson_updates_2018}, but it seems
unlikely that we would be able to generate such dynamics with the
models presented in this article. Additional nonlinearities would
be needed to create intrinsically variable and chaotic dynamics.

With these assumptions in mind, we have found that first, temporally
forced Lotka-Volterra dynamics cannot sustain any diversity with our
phytoplankton-based set of parameters, unless the structure is geared
to include either a storage effect or a strong self-regulation. Although
this absence of diversity-enhancing effect of ``pure'' environmental
variation has already been stated by other authors \citep{chesson_roles_1997,barabas_community_2012,fox_intermediate_2013,scranton_coexistence_2016},
this is not always intuitive \citep{fox_intermediate_2013}, so we
feel compelled to stress it once more: temporal variation in growth
rate alone cannot help coexistence within competitive communities.
A nice point made by \citet{scranton_coexistence_2016} was that a
built-in storage effect in a forced Lotka-Volterra model, parameterized
for phytoplankton communities, could lead to a reasonable degree of
coexistence. Our investigation reproduced these results, using the
random noise considered by \citet{scranton_coexistence_2016}. However,
an arguably more lifelike noisy and seasonal temperature forcing considerably
lessened the richness of the community after 5000 years, decreasing
from 15 to 4 species on average. Even imagining that groups represented
here are genera or classes rather than species, this is a fairly low
diversity for a phytoplankton-like community (see e.g., Chapter 1
in \citealp{reynolds2006ecology}). This suggests that the storage
effect may not, on its own, be sufficient to maintain species-rich
communities (e.g., dozens to hundreds of species). We have therefore
sought out whether a stronger self-regulation could maintain a higher
diversity, using field-based intra- vs intergroup (species or genera)
competition strength ratio \citep{barraquand2018coastal}, where the
intragroup density-dependence was estimated 10 times stronger. Implementing
such strong self-regulation, in the forced Lotka-Volterra models that
we considered, produced a higher level of diversity than the storage
effect (almost double). Of course, the result is somehow contingent
upon the strength of self-regulation. Our estimates are stronger
than what was found in perennial plants \citep{adler_coexistence_2010},
where interspecific competition was suggested 4 or 5 times stronger
than intraspecific. Still, the widespread effects of natural enemies
in phytoplankton (zooplankton, parasites) may contribute to an increase
in the self-regulation strength \citep{barraquand2018coastal,chesson_updates_2018}
relative to other systems, hence we believe that 10 times stronger
intraspecific competition constitutes a reasonable order of magnitude.

However, such strong self-regulation was still insufficient to maintain
the whole community diversity (60 species) by itself, especially when
the seasonal forcing was considered (always decreasing species richness).
The diversity within clumps of similar values of thermal optima was
considerably decreased once seasonality was implemented. This diversity
reduction occurs because within a season, the signal autocorrelation
gives long, contiguous time intervals to the best competitor to exclude
its less adapted competitors. This makes the results likely to hold
not only for seasonal environments, but more generally for autocorrelated
ones above the daily scale, i.e., ``red'' noise. In contrast, the
random noise scenario -- which can be considered white noise above
the daily temporal scale -- generates large temperature shifts more
frequently, and thereby forbids such competitive exclusion. In a seasonal
setting, a species with the highest long-term (arithmetically) averaged
growth rate may not be the best competitor, and can disappear as a
result of a strong competition from both low- and high-temperature
tolerant species. This holds with or without a storage effect.

Our results may appear at odds with recent proposals that seasonal
forcing in itself would help maintain diversity \citep{sakavara_lumpy_2018}.
However, we compared the effect of seasonal forcing to that of other
forcing signals while controlling for total variance. Thus, the contrast
between our results and those of \citet{sakavara_lumpy_2018} may
be due to the role of forcing variance over time: we compare scenarios
under a constant total variance, much like what is done when examining
the effect of noise color on population and community dynamics \citep{jiang_temperature_2007,ruokolainen_ecological_2009}.
Thinking in terms of signal spectrum, while seasonality may maintain
slightly more diversity than no forcing at all if a storage effect
is present, the reddening of the environmental noise due to such seasonality
reduces coexistence. This result may be contingent upon the correlated
positive responses of the species growth rate to increases in the
environmental variable \citep[and references therein]{ruokolainen_ecological_2009}.

The biomass-trait relationship was affected more marginally by the
type of forcing signal. The storage effect alone begot several clumps
along the trait space (as observed by \citealp{scranton_coexistence_2016}).
The seasonality that we added to the temperature signal led to more
distant clumps on the trait axis, with less species per clump. Conversely,
strong self-regulatory mechanisms alone led to relatively uniform
biomass distributions, with species forming a single large cluster,
which covered a fraction of the initial trait space. Therefore, the
shape of the distribution was mostly affected by the coexistence mechanism
at work while the average trait value was modified by the type of
environmental forcing, even though the mean value of the environmental
signal did not change. However, when both strong-self regulation and
the storage effect were at play, the biomass-trait distribution could
either be unimodal or multimodal depending on the type of noise driving
the community dynamics (random or seasonal, respectively). This implies
that the mere observation of multimodality in a thermal preference
trait-biomass distribution is not a proof of a storage effect, or
conversely, the proof of the influence of a seasonal environment.

The identification of multiple modes in biomass-trait distributions
is relatively recent \citep{segura_competition_2013,loranger_what_2018,dandrea_translucent_2018,dandrea_generalizing_2019},
so we recommend to interpret them with caution to avoid over-generalization.
\citet{barabas_emergent_2013} convincingly argued that multimodality
could arise from the demographic stochasticity of a single model run.
However, with several locations - or in a theoretical context as done
here - one could average across locations. There are additional reasons
to be cautious: the occurrence of clustering is very sensitive to
the shape of the competition kernel; small differences in shape can
shift the distribution towards either clustered or uniform \citep{pigolotti_how_2010}.
We therefore view clustering on the thermal preference trait axis
as an interesting clue suggesting to look for a storage effect, rather
than any definite proof that the storage effect is at work. Finally,
we recall that we focus on a trait (thermal optimum) which clearly
interacts with the environment: clustering may emerge on another trait
axis, such as size, which typically affects the competition coefficient,
without having any relationship to the storage effect \citep{segura_emergent_2011,segura_competition_2013,dandrea_translucent_2018,dandrea_generalizing_2019}.

In our previous empirical study of phytoplankton dynamics \citep{barraquand2018coastal},
we did not find any storage effect. This does not mean that it could
not be observed in other planktonic systems: we studied a coastal
ecosystem and focused a specific fraction of phytoplankton, relatively
large diatoms and dinoflagellates. However, given the consequences
of the storage effect for species richness and composition presented
here, we are skeptical that the storage effect could, by itself, fully
explain phytoplankton diversity at any location. Our results suggest
that in phytoplankton-like seasonal environments, empirically-tuned
self-regulation produces much more diversity than the storage effect,
when both are considered in isolation. The storage effect may therefore
help phytoplankton diversity maintenance, but only when combined to
other mechanisms. This is all the more likely that in our models,
the combination storage effect + strong self-regulation is non-additive:
the cases where both self-regulation and the storage effect were present
showed more diversity than generated by any mechanism on its own.

The above results suggest the very exciting idea that multiple coexistence
mechanisms might combine superadditively to determine the richness
of the community, thus helping us to better understand the astounding
diversity of primary producers. This logic could, in principle, be
extended to mechanisms that we have not considered here (e.g., spatial
structure, specialized natural enemies, that could be as important
here for plankton as they are for tropical trees, see \citealp{bagchi_pathogens_2014,comita_testing_2014,barraquand2018coastal}).
Superadditivity, i.e. the positive effect of interactions between
mechanisms can be measured either on community diversity, as we did
here, or on the invasion growth rates \citep{ellner_expanded_2019}.
Using the latter metric, previous research has however demonstrated
that generalist seed predation could weaken the storage effect \citep{kuang_coexistence_2009,kuang_interacting_2010}
thus different mechanisms might not always combine superadditively
as we found here. That said, superadditivity has been found in some
cases, i.e., pathogens could enhance the storage effect and broaden
the conditions in which species could coexist \citep{mordecai_pathogen_2015}.
Better explaining plant or microbial diversity would then not be about
selecting the best unique mechanism susceptible to explain the observed
diversity, but rather better combining those mechanisms together.
This may obviously be an annoyance for those who like to sharpen Occam's
razor, but it clearly holds opportunities for theoreticians wishing
to investigate synergies between coexistence mechanisms in highly
diverse communities. Aside from the synergies between predator diversity-enhancing
effects, strong self-regulation through various means and the storage
effect (on the temporal axis), one obvious follow-up of this research
would be interactions with spatial structure. Spatial structure occurs
both endogeneously, through spatially restricted movements and interactions,
and exogeneously, through spatial variation in environmental covariates
\citep{bolker_combining_2003}. Numerous studies \citep[e.g.,][]{bolker_spatial_1999,murrell_2002}
have shown that spatially restricted movements and interactions -
very small-scale spatial structure - can help coexistence, which we
believe would be especially important for phytoplankton since many
species form colonies (\citealp{reynolds2006ecology}; see discussion
in \citealp{barraquand2018coastal}). Moreover, although temperature
is usually relatively spatially homogeneous over space, other drivers
(e.g., rainfall, pH in terrestrial ecosystems; salinity in aquatic
ones) may exhibit spatial variation which is a main factor for coexistence
\citep{snyder_when_2008}. The odds that different (resource) niches,
natural enemies, spatial limits to competition and temporal niche
partitioning all interact to promote the very high-dimensional coexistence
observed in the field seem much higher than for any of those mechanisms
alone. Whether the diversity-enhancing effects of these mechanisms
combine subadditively (as in \citealp{kuang_interacting_2010}) or
superadditively like here is therefore worthy of further research.

\section*{Acknowledgements}

We thank Alix Sauve for thoughtful comments and some bibliographic
references. We are very grateful to Gy\"orgy Barab\'as and an anonymous
referee for their constructive feedback. This study was supported
by the French ANR through LabEx COTE (ANR-10-LABX-45).

\newpage
\section*{Electronic Supplementary Material}
\subsection*{Biomass-trait distributions} %Was before supplementary figures
\setcounter{figure}{0}
\renewcommand\thefigure{A\arabic{figure}} 
\begin{figure}[ht!]
\begin{centering}
\includegraphics[width=0.95\textwidth]{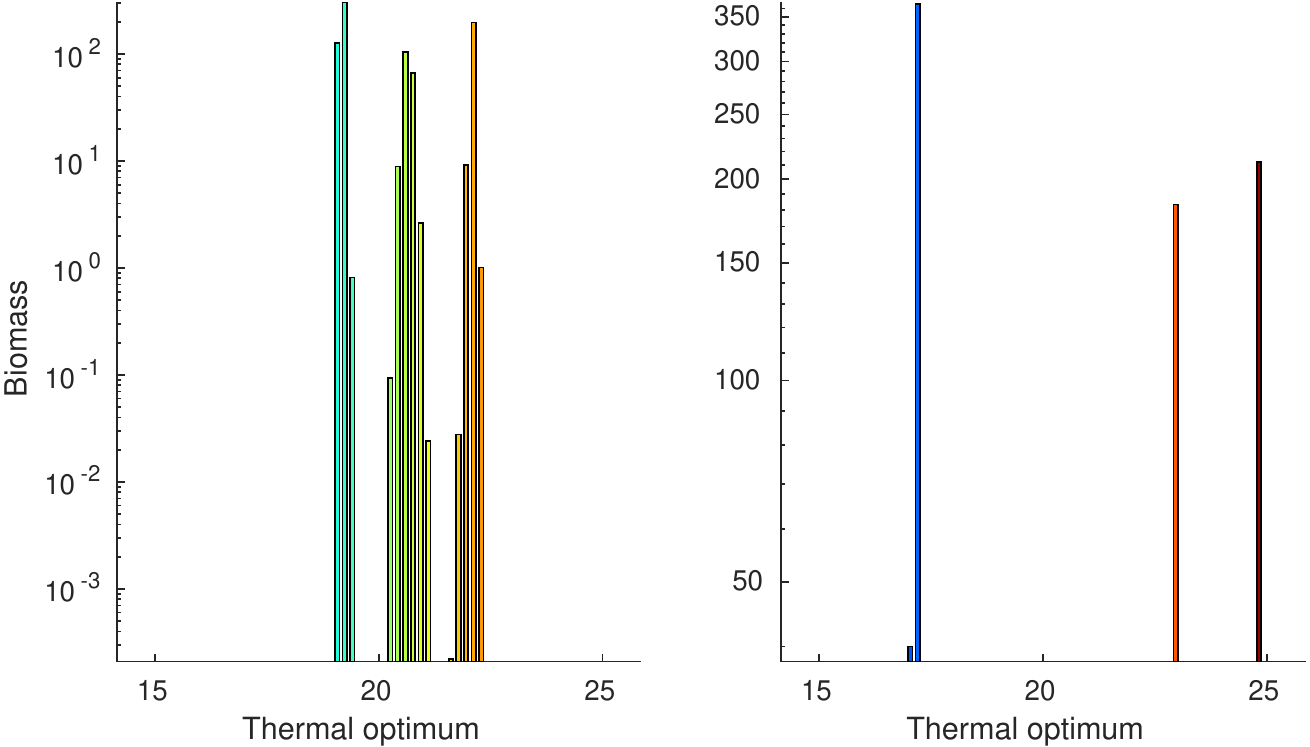}
\par\end{centering}
\caption{Temporal mean of biomass as a function of the thermal optimum defining
each species. The temporal means are computed over the last 200 years
of a simulation spanning 5000 years. We considered both a random
(left) and a seasonal signal for the temperature (right). The coexistence mechanism
implemented is the storage effect, and the intra and interspecific competition coefficients are equal. This
simulation is the one described in Fig.~1
in the main text\label{fig:Appendix_mean_biomass_iter2}. 99 other simulations
have been performed to produce the main text results in Figs. 2-4. }
\end{figure}

\begin{figure}[!ht]
\begin{centering}
\includegraphics[width=0.95\textwidth]{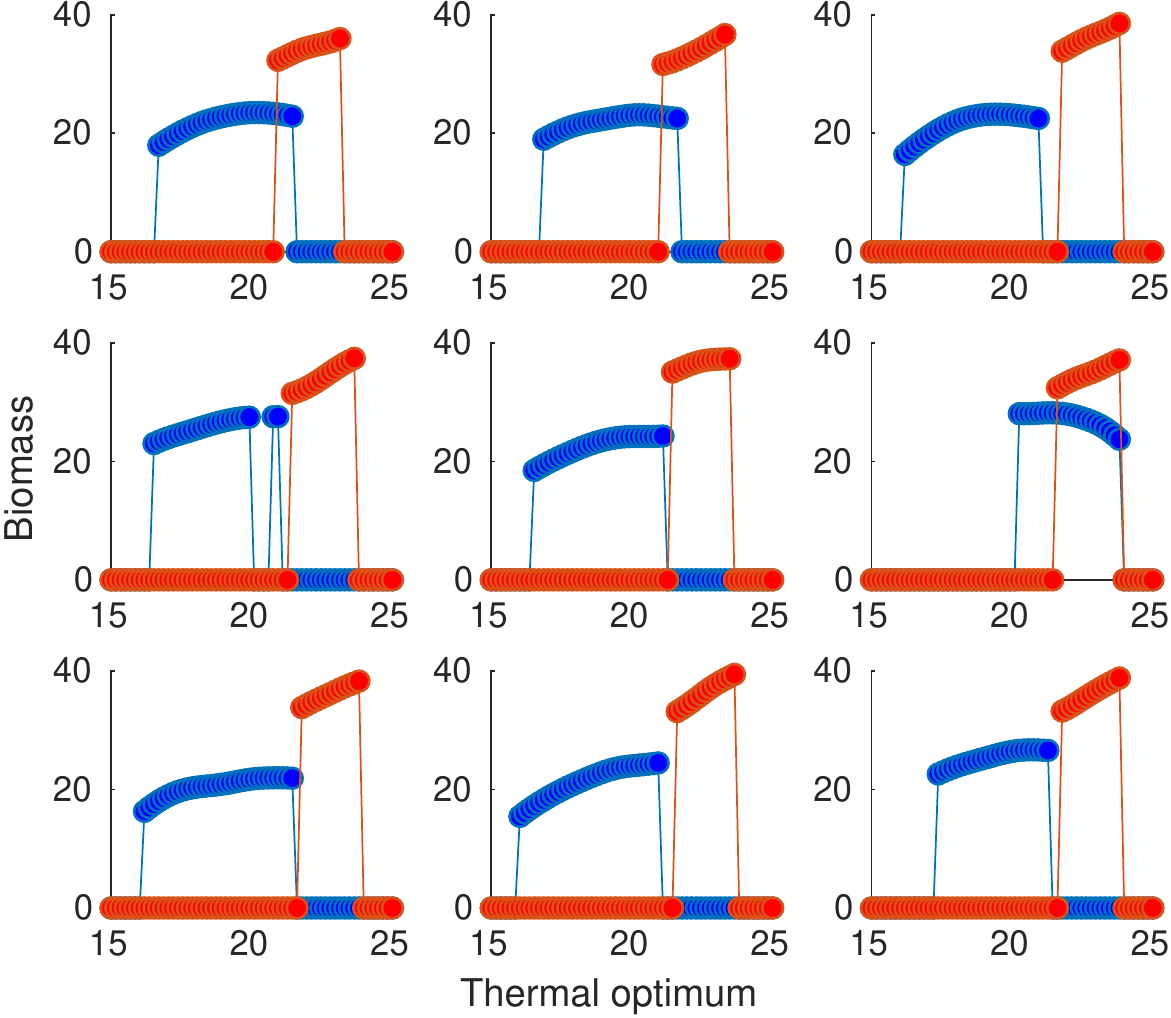}
\par\end{centering}
\caption{Temporal mean biomass distribution, computed over the last 200 years,
for 9 representative simulations, as a function of the thermal optimum
of the species. These simulations are done without storage effect
but with a strong self-regulation. Temperature is either a seasonal
signal (red) or a random noise (blue). The distribution induced
by a random noise overlaps the one obtained with a seasonal noise in only 2 simulations out of 100, hence the 2 signals
lead in general to non-overlapping biomass distributions on the
trait axis. \label{fig:Not_representative_behaviour} }
\end{figure}

\newpage
\subsection*{Variation in mortality rates}

To test the robustness of our conclusions while accelerating convergence, we conducted the same analyses with a species-specific mortality rate. For each set of simulations, covering 4 different competition scenarios and 2 types of environmental forcing, the mortality rate was drawn from a uniform distribution between 14.9 and 15.1 year$^{-1}$ so that we only changed the variability, but not the mean, of this parameter. 

\begin{figure}[!ht]
\begin{centering}
\includegraphics[width=0.95\textwidth]{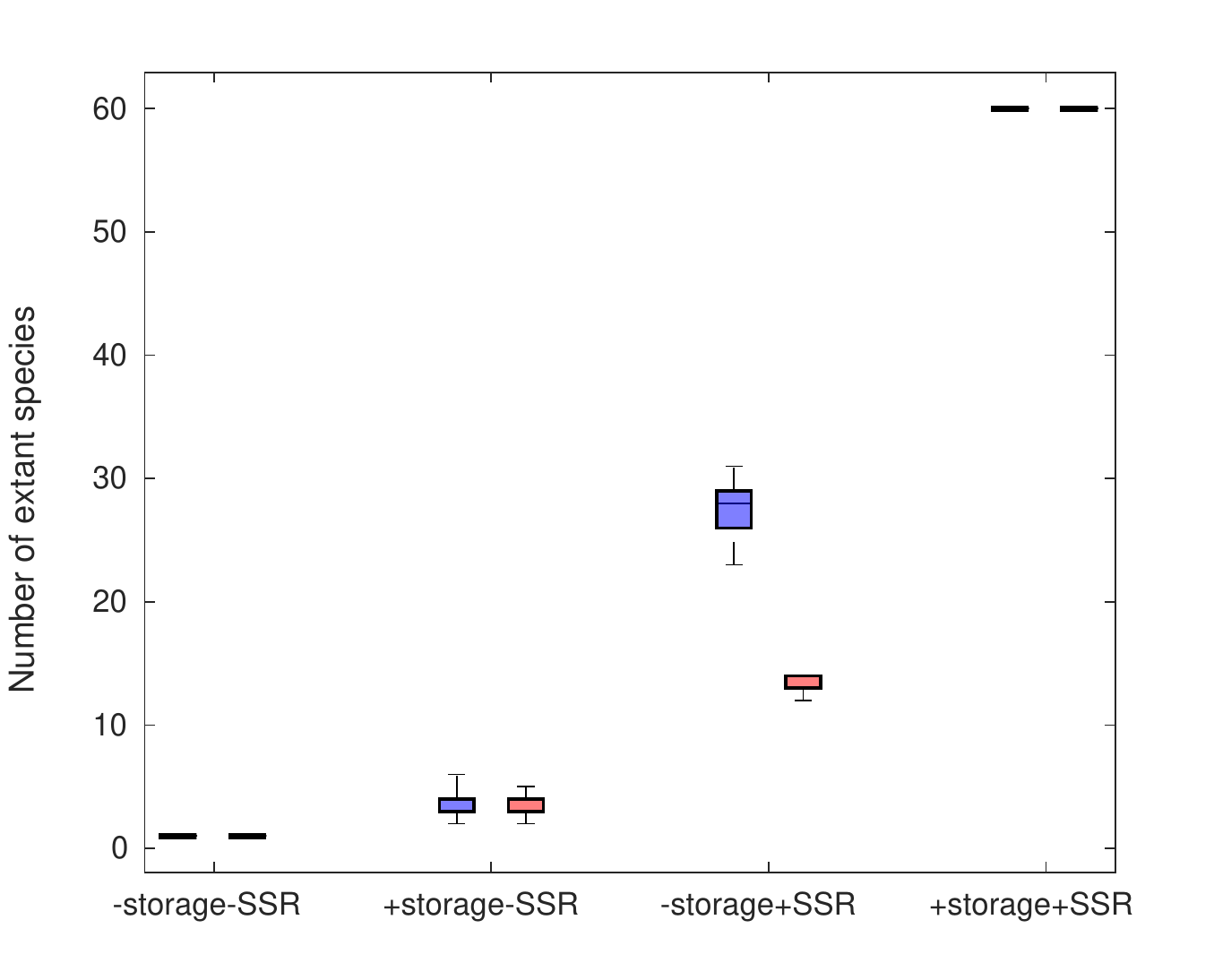}
\par\end{centering}
\caption{Number of species still present at the end of 100 simulations (5000 years each) with a variable mortality,
initialized with 60 species, with a random forcing signal (blue) or a seasonal noise (red). The
signs + or -storage refer to presence or absence of the storage effect, respectively; + / - SSR,
presence or absence of Strong Self-Regulation, respectively. Community compositions are stable
in the cases -storage-SSR and +storage+SSR, for which 1 or 60 species are still present at the
end of all simulations. Due to low variance, the whiskers here represent min and
max rather than 1.5 interquartile range.\label{fig:Nb_extant_morta_variable}}
\end{figure}

\begin{figure}[!ht]
\begin{centering}
\includegraphics[width=0.95\textwidth]{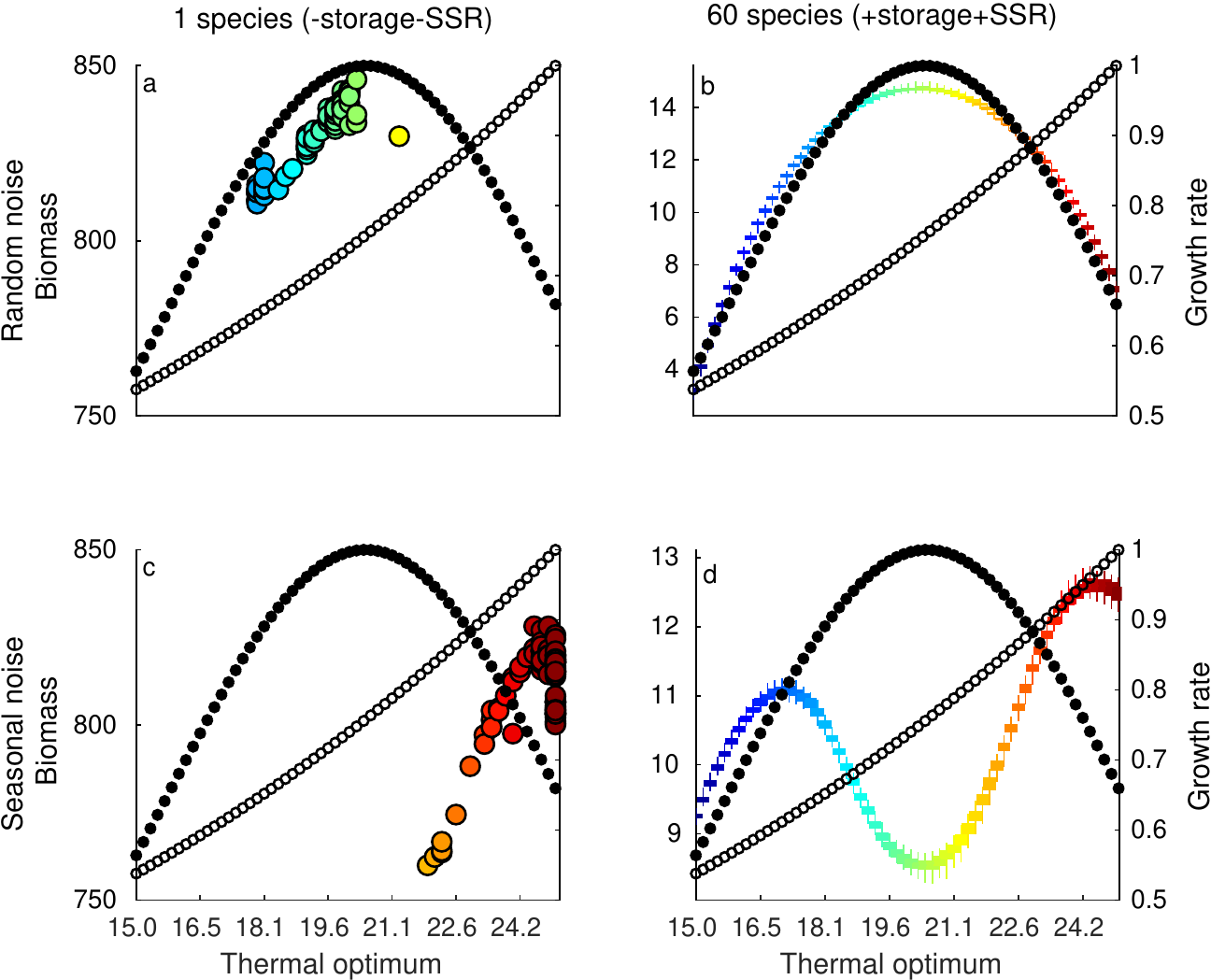}
\par\end{centering}
\caption{Mean biomass distribution over the last 200 years for 100 simulations, as a function of the species thermal optima. 
We consider here a variable mortality between species. Here we consider the
two stable-composition cases and two types of forcing signal. On the
left column, simulations without storage effect nor strong self-regulation
are presented. Only one species is present at the end of the simulations
and its mean value is represented by one large colored circle per
simulation. There can be several circles for the same species, corresponding
to multiple simulations ending with this species alone. On the right
column, simulations with storage effect and strong self-regulation
are represented. All species are present at the end of the simulations
and small boxplots present the variation in the temporal average of
biomass with a given trait, for 100 simulations. The forcing signal
is either a random (top) or a seasonal noise (bottom). Each species
is identified by its thermal optimum through its color code. Scaled
(divided by maximum) average and maximum growth rates are shown as
small filled and open circles, respectively, and are indexed on the
right y-axis.\label{fig:Fig3_morta_variable}}
\end{figure}

\begin{figure}[!ht]
\begin{centering}
\includegraphics[width=0.95\textwidth]{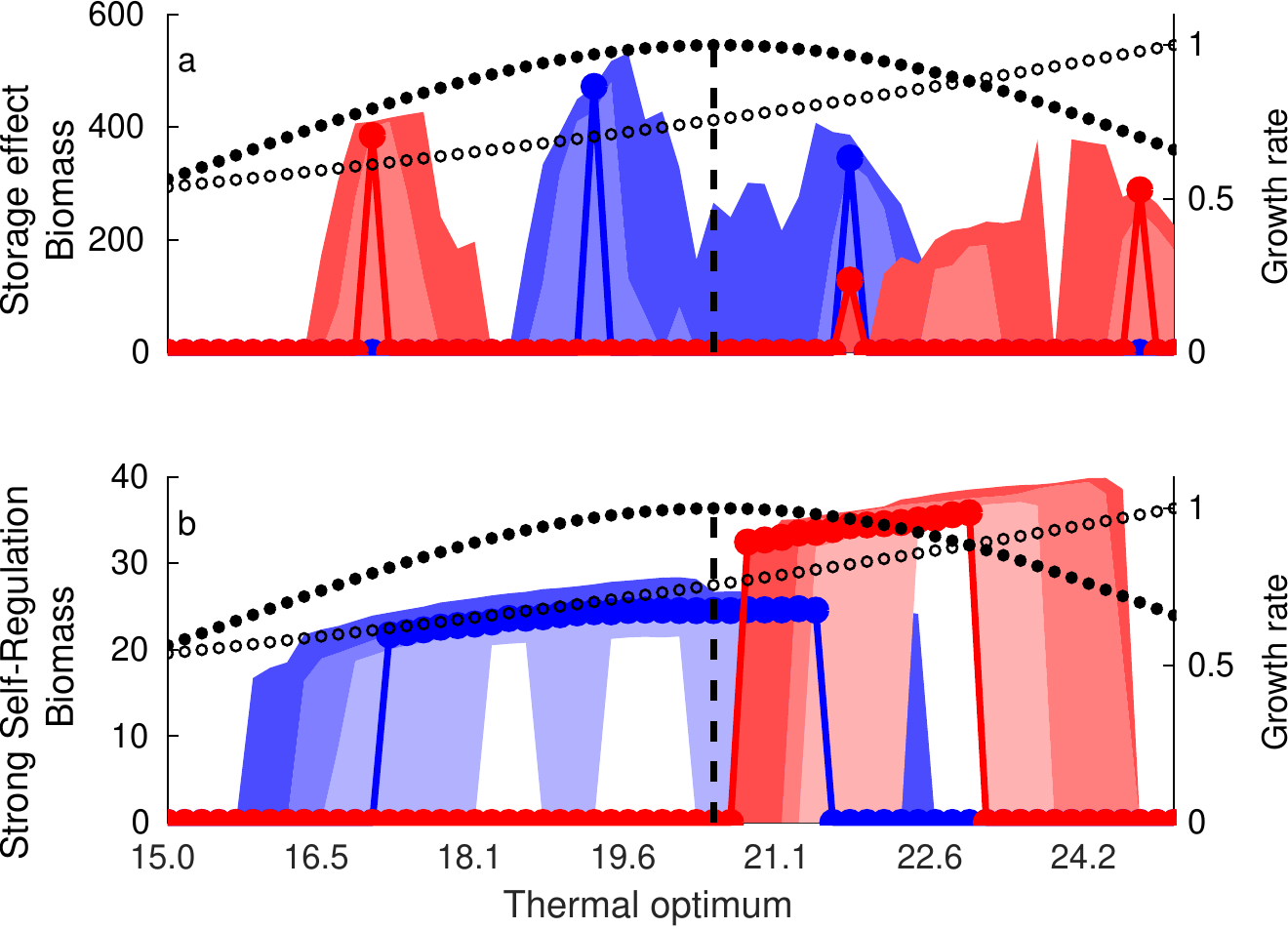}
\par\end{centering}
\caption{Mean biomass distribution over the last 200 years for 100 simulations as a function of thermal optima. 
We consider here a variable mortality between species. The two cases considered are (a) with storage effect and equal competitive
strengths and (b) without storage effect, with strong self-regulation. The forcing signal is either a random (in blue) or a seasonal
noise (in red). Shades of the same color correspond to the 50th, 90th and 100th percentiles of the distributions while colored lines correspond
to one representative simulation. Scaled (divided by maximum) average (whose maximum is indicated by the dashed line) and maximum growth
rates are shown as filled and open and circles, respectively, and indexed on the right y-axis.\label{fig:Fig4_morta_variable}}
\end{figure}

The main results of our analyses were not altered by this modification (Fig.~\ref{fig:Nb_extant_morta_variable}). The absence of coexistence mechanisms led to competitive exclusion of all species but one and the presence of both coexistence mechanisms maintained all species (Fig.~\ref{fig:Fig3_morta_variable}). Strong self-regulation on its own maintained between 23 and 31 species (vs 20 to 32 in the case of constant mortality) with a random noise, and between 12 and 14 (vs 12 to 15 with a constant mortality) with a seasonal noise. The storage effect alone also led to similar results with a seasonal noise with regards to the richness of the community. As shown on Fig.~\ref{fig:Fig4_morta_variable}, biomass-trait distributions remained qualitatively similar (multimodality with the storage effect and uniform distribution with a strong self-regulation, with different partitioning on the trait axis depending on the type of noise). The only case which led to slightly different results was the combination of a storage effect and random noise. In this case, the final number of species in the community ranged from 2 to 6, with nearly 50\% of the simulations ending with 3 species only. Richness is therefore approximately 4 times lower than what was obtained with a constant mortality. The remaining species had approximately the same positions on the trait axis as in the case of a constant mortality. 

Competitive exclusion within clumps was therefore accelerated by the variation in mortality. This can be explained by the departure from neutral dynamics which, in the absence of immigration, led to the exclusion of species being even marginally more vulnerable than the others within their clumps. We can assume that the results obtained with a variable mortality mimic the ones that could be obtained with a constant mortality but for much longer runtimes, after the longest transients. 

\end{document}